\documentclass[aip,apl,10pt,twocolumn,superscriptaddress,10pt]{revtex4-1}
\usepackage{amssymb,amsmath,amsfonts,xcolor,graphicx}

\begin{document}

\title{\Large Circuit electromechanics with single photon strong coupling}

\author{Zheng-Yuan Xue} \email{zyxue@scnu.edu.cn} 
\affiliation{Guangdong Provincial Key Laboratory of Quantum Engineering and Quantum Materials, and School of Physics and Telecommunication Engineering, South China Normal University, Guangzhou 510006, China}

\author{Li-Na Yang}
\affiliation{Guangdong Provincial Key Laboratory of Quantum Engineering and Quantum Materials, and School of Physics and Telecommunication Engineering, South China Normal University, Guangzhou 510006, China}

\author{Jian Zhou} \email{jianzhou8627@163.com}
\affiliation{Department of Electronic Communication Engineering, Anhui Xinhua University, Hefei, 230088, China}
\affiliation{Guangdong Provincial Key Laboratory of Quantum Engineering and Quantum Materials, and School of Physics and Telecommunication Engineering, South China Normal University, Guangzhou 510006, China}

\date{\today}

\begin{abstract}
In circuit electromechanics, the coupling strength is usually very small. Here, replacing the capacitor in circuit electromechanics by a superconducting flux qubit, we show that the coupling among the qubit and the two resonators can induce effective electromechanical coupling which can attain the strong coupling regime at the single photon level with feasible experimental parameters.  We use dispersive couplings among two resonators and the qubit while the qubit is also driven by an external classical field. These couplings form a three-wave mixing configuration among the three elements where the  qubit degree of freedom can be adiabatically eliminated, and thus results in the enhanced coupling between the two resonators. Therefore, our work constitutes the first step towards studying quantum nonlinear effect in circuit electromechanics.
\end{abstract}


\maketitle

Circuit electromechanics, a microwave version of cavity optomechanics \cite{om}, explores the interaction between the electromagnetic radiation from a transmission line resonator and the motion of a mechanical resonator  \cite{em1,linear1,em2,strong,em3}. Besides testing the quantum theory in macroscopic systems, it also finds other interesting applications in modern physics. With rapid experimental progress, ground state cooling of a mechanical resonator has already been achieved in the linear regime \cite{linear1,linear2}, where the effective interaction is substantially enhanced by a strong externally driven cavity mode. In this regime, the ordinary thermal and vacuum noises lead this kind of  system to a Gaussian steady state.

The single-photon coupling is non-linear, and thus can be used to observe non-Gaussian physics in these systems. To this end, one needs to achieve the single-photon strong coupling regime, where the coupling strength,  due to the single-photon radiation pressure, exceeds the cavity decay and the mechanical damping. Nevertheless, in typical electromechanical setups, the coupling strength tends to be very small \cite{strong}, and thus needs to be greatly increased. This single-photon coupling strength is shown to be substantially enhanced in the electromechanical system by replacing the coupling capacitor by a superconducting qubit, where the strong coupling between the mechanical and microwave resonators can be induced due to the intrinsic nonlinearity of the qubit \cite{enhance1,en1,en2,enhance2,enhance3,en3}. Similarly, this enhancement can also be obtained with produced nonlinearity for the cavity or mechanical modes \cite{mode1,mode2,mode3}.

Superconducting circuits are very flexible, and thus have very interesting applications in modern quantum technology \cite{t1,t2,t3,cqed}.  Recently, there has been increasing interest in using superconducting flux qubits,  as artificial atoms with three levels, to investigate many atomic and quantum optical phenomena, in particularly, as  a $\Delta$ configuration atom that is absent in nature \cite{3l,3l1}. Experimentally,  the frequency up-conversion of a microwave photon in this $\Delta$ configuration  flux qubit was recently demonstrated \cite{3lexp}. Meanwhile, electrical \cite{c1,fluxcavity,c2,c3} and mechanical \cite{m1} resonators are demonstrated to be able to coherently couple to a superconducting flux qubit in the strong coupling regime.

Here, we propose to induce a strong electromechanical coupling by a superconducting flux qubit with three levels in a $\Delta$ configuration. Specifically, in our scheme, we use dispersive couplings among two resonators and the qubit while the qubit is also driven by an external classical field. These couplings form a three-wave mixing configuration among the three elements where the  qubit degree of freedom can be adiabatically eliminated, and thus results in the enhanced coupling between the two resonators. We show that the effective strong coupling at the single photon level can be induced with feasible experimental parameters. Therefore, the nonlinear effect in this system may be further explored.


The proposed setup is shown in Fig. 1a, where a superconducting flux qubit couples dispersively to both the  electric and mechanical resonators. The flux qubit consisting of three Josephson junctions in a loop, where one of the junctions is smaller by a factor $0.5<\alpha<1$ than the other two junctions.  For this qubit, the first two eigenstates are localized within their respective potential wells, while the third one is delocalized. When the loop is biased by a magnetic flux $\Phi$ that is slightly deviated from its symmetric point, i.e., $\Phi\neq \Phi_0/2$ with $\Phi_0=h/(2e)$ being the flux quanta, the symmetry of the qubit's potential wells is broken \cite{3l1} and the quantum states within the two wells can have a nonzero dipole transition element. Therefore, it can serve as a $\Delta$ type artificial atom with a cyclic transition configuration. In this case, there is no selection rule governing the state transitions. The only way to implement coupling on a certain transition is to use frequency selection, i.e., the frequency is close to the target transition while far away from the others.

\begin{figure}[tb]
\begin{center}
\includegraphics[width=0.99\columnwidth]{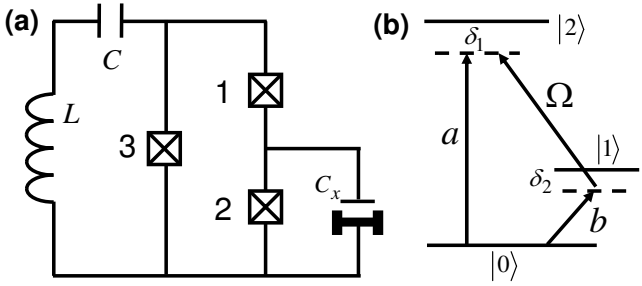}
\end{center}
\caption{The proposed setup for our proposal.  (a) The circuit representation of a superconducting flux qubit couples both to the electrical and mechanical resonators. (b) The coupling structure in terms of flux qubit energy levels.}
\label{f1}
\end{figure}

We label the three lowest energy levels as $|k\rangle$ with $k \in \{0, 1, 2\}$, and the qubit Hamiltonian can be written as $H_q=\sum_{k=0}^2 \omega_{k} |k\rangle\langle k|$,  assuming $\hbar=1$ hereafter. The energy difference between qubit eigenstates is denoted by $\omega_{k,l}=\omega_{k}-\omega_{l}$ with $k>l$ to assure $\omega_{k,l}>0$, which can be modulated via the external control dc magnetic flux. For typical qubit design, $\omega_{k,l}$ can be made to be within the range of 2$\pi\times$ [0, 10] GHz. To make all the cyclic transitions  addressable, we modulate the magnetic bias to make the  eigenstates have large anharmonicity, i.e., $\omega_{2,1}=2\pi\times 4.28$ GHz and $\omega_{1,0}=2\pi\times 1.72$ GHz. This can be achieved by a proper choice of the qubit parameters, or making the energy level of $|1\rangle$ to be tunable by replacing the third junction of the flux qubit with a superconducting quantum interference device and modulating its enclosed magnetic flux \cite{tune1,tune2}.

In our implementation, the flux qubit is also inductively driven by an external microwave field, which dispersively couples to the transition of $|1\rangle\leftrightarrow|2\rangle$. Microwave signals can be applied to the qubit via an on-chip antenna  \cite{3lexp}. Typically, this can be modeled as
\begin{eqnarray}
H_1=\Omega\exp(i\omega t) \sigma_{1,2} +\text{H.c.},
\end{eqnarray}
where $\sigma_{l,k}=|l\rangle\langle k|$ with $l<k$ are the downgrade operators of the qubit states while its hermitian conjugates are the upgrade operators $\sigma_{l,k}^\dagger=|k\rangle\langle l|$, $\Omega$ and $\omega=2\pi\times 4.28$ GHz are the amplitude and frequency of the driven field, respectively.

The electric resonator can be a lumped-element harmonic oscillator or a superconducting transmission line resonator \cite{cqed}, which forms a 1D cavity with frequency $\sim 2\pi\times [1, 10]$ GHz, decay rate $\kappa_a$ of several kHz, and size on the millimeter scale. Here, for demonstration purpose, we set $\kappa_a= 2\pi\times 5$ kHz. The superconducting flux qubit is fabricated close to the magnetic antinode of the  the electric resonator  and coupled to the resonator via the mutual inductance, where the coupling strength is proportional to the total inductance shared between the resonator and the qubit, and thus strong coupling can be attained while minimizing sensitivity to the flux noise \cite{fluxcavity}. A similar architecture may be able to investigate ultra-strong coupling between the flux qubit and the 1D cavity \cite{ultrastrong,t4}.
In addition, we assume that the frequency of the first electric resonator mode $\omega_a=2\pi\times 5.68$ GHz is close to $\omega_{2,0}=2\pi\times 6$ GHz, and thus it is the only involved mode of the resonator.

The involved mode of the electric resonator can be modeled as $H_e= \omega_a a^\dagger a$. In the qubit eigen  basis, this dispersive coupling can be described by \cite{ultrastrong}
\begin{eqnarray}  \label{heq}
H_2 = g_1 a^\dagger \sigma_{0,2}+\text{H.c.},
\end{eqnarray}
where $g_1$, $a$ and $a^\dagger$ are the coupling strength, annihilation and creation operators of the the electric resonator, respectively. In a recent experiment with a transmission line resonator \cite{fluxcavity}, this kind of coupling strength is demonstrated to be $\sim 2\pi\times 100$ MHz. Here, we choose $g_1=2\pi\times 40$ MHz.

Meanwhile, the flux qubit is also capacitively coupled to a GHz mechanical resonator \cite{m1}.  With current experimental techniques,  at low temperatures, a nanocrystalline diamond mechanical resonator can be built \cite{nm1} with quality factors $Q \sim 10^4$ and fundamental mode frequency of $\omega_b\approx 2\pi\times 1.4$ GHz, which corresponds to decay rate $\kappa_b \sim 2\pi \times 0.1$ MHz. With the  level spacing of $\omega_{1,0}=2\pi\times 1.72$ GHz, it is possible to make the fundamental vibrational mode of the mechanical resonator $\omega_b$ only relevant to the wanted transition of $|0\rangle\leftrightarrow|1\rangle$ with large detuning. The mechanical resonator can be modeled as $H_m= \omega_b b^\dagger b$.  In the qubit eigen  basis, this coupling is described by
\begin{eqnarray}  \label{hmq}
H_3 = g_2 b^\dagger \sigma_{0,1} +\text{H.c.},
\end{eqnarray}
where $g_2$, $b$ and $b^\dagger$  are the coupling strength, annihilation and creation operators of the mechanical resonator, respectively.  In the experiment of Ref.  \cite{m1}, the coupling strength between the flux qubit and the mechanical resonator is about 2\% of the mechanical fundamental frequency. Assuming a little better than this efficiency, for our mechanical resonator with $\omega_b\approx 2\pi\times 1.4$ GHz, we choose $g_2 = 2\pi\times 40$ MHz.

As show in Fig. 1b, due to the fact that $\delta_1=\omega_{2,0}-\omega_a=\delta_2=\omega_{1,0}-\omega_b=2\pi\times 320$ MHz, all the above three couplings can only perturbatively couple to the flux qubit. Meanwhile, they are in three-photon resonance, i.e., $\omega_a=\omega_b+\omega$. In this way, assuming the qubit is initially prepared in the ground state, we only virtually  excite the higher energy levels, and thus the limited lifetime of them will not introduce significant noise into the quantum dynamics of the coupled system. Moreover, this will greatly simplify the quantum dynamics of the combined coupled system as we can adiabatically eliminate the flux qubit's degree of freedom. To be more specific, when $\delta_1\gg g_1$, $\delta_2\gg g_2$, and $\delta_1\delta_2\gg \Omega^2$, the adiabatic elimination  \cite{eliminate} of the flux qubit state can be achieved. That is, in the interaction with respective to $H_0=H_e+H_m+H_q$, the interaction Hamiltonian $H_{int}=H_1+H_2+H_3$ will reduce to the effective coupling between the electric and mechanical resonators as
\begin{eqnarray}
H_{em}\approx \langle 0|H_{int}|0\rangle \approx \lambda(a^\dagger b+ ab^\dagger),
\end{eqnarray}
where $\lambda\approx\Omega g_1g_2/(\delta_1\delta_2)$, and  we have omitted a small modification of the resonant frequencies for the two resonators $\lambda_i\approx\delta_ig_i^2/(\delta_1\delta_2)$ with $i\in \{1,2\}$, which can easily be compensated by tuning the resonators.

To ensure the adiabatic elimination, the amplitude of the driven field $\Omega$ should be $\Omega\ll \sqrt{\delta_1\delta_2} = 2\pi \times 320$ MHz. Therefore, it can be chosen as  $\Omega=2\pi \times 64$ MHz, which leads to $\lambda=2\pi \times 1$ MHz. It is obvious that this reduced coupling strength is much larger than the decay of both resonators, and thus it is within the strong coupling regime. Moreover, this coupling is induced by the single photon interactions in Eqs. (\ref{heq}) and (\ref{hmq}), instead of using many photons in conventional optomechanical and electromechanical experiments  \cite{em1,linear1,em2,strong,em3}, and thus the scheme presented here is well in the strong coupling regime at the single photon  level. Alternatively, to decrease the technical difficulties of hybridizing elements fabricated by different materials, one may prefer to use an aluminium based  mechanical resonator \cite{nm2}. In this case, to obtain large coupling strength $g_2$, one should choose  mechanical resonators with frequencies at the GHz level. However, such resonators usually do not have high enough quality factors. If one chooses resonators with frequencies on the order of 0.1 GHz (with additional voltage drive to address the prescribed $|0\rangle\leftrightarrow|1\rangle$ transition), they can possess high enough quality factors, but the coupling strength should be much smaller. Fortunately, assuming the overall same performance, the resulting electromechanical coupling is about  $\lambda/10$, which can still be in the strong coupling regime.

\begin{figure}[tb]
\begin{center}
\includegraphics[width=0.9\columnwidth]{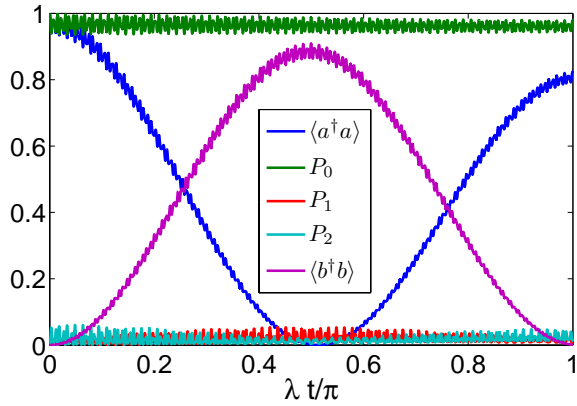}
\end{center}
\caption{Transfer of photon and phonon states and the population of the qubits states. $P_k$ denotes the population of the level $|k\rangle$.}
\label{f1}
\end{figure}

The performance of  our implementation can be evaluated by considering the influence of dissipation using the  quantum master equation of
\begin{eqnarray}  \label{me}
\dot\rho &=& i[\rho, H_{int}] +\frac{\kappa_a}{2}  \mathcal{L}(a)  +\frac{\kappa_b}{2}  \mathcal{L}(b)  +\frac{\Gamma_1}{2} \left[
 \mathcal{L}(\sigma_{0,1}) + \mathcal{L}(\sigma^\text{z}_{0,1})\right] \notag\\
&&+ \frac{\Gamma_2}{2} \left[
 \mathcal{L}(\sigma_{0,2}) + \mathcal{L}(\sigma^\text{z}_{0,2})
+\mathcal{L}(\sigma_{1,2}) + \mathcal{L}(\sigma^\text{z}_{1,2})\right],
\end{eqnarray}
where $\rho$  is the density matrix of the considered total hybrid system, $\mathcal{L}(A)=2A\rho A^\dagger-A^\dagger A \rho -\rho A^\dagger A$ is the Lindblad operator, and $\sigma^\text{z}_{l,k}=|k\rangle\langle k|-|l\rangle\langle l|$ with $k<l$. As $|0\rangle$ and $|1\rangle$ are localized states in the qubit's two potential wells, they are usually selected as the qubit states for flux qubit as they have relatively long coherent times. For our demonstration purpose, we choose the decay and dephasing rates of these qubit states to be the same as $\Gamma_1=2\pi \times 10$ kHz.  For the delocalized excited state $|2\rangle$, its lifetime is usually short comparing with the localized qubit states. Here, we assume its decay and dephasing with respect to both the qubits states $|0\rangle$ and $|1\rangle$ are the same. Meanwhile, we also assume its decay and dephasing are the same as $\Gamma_2=2\pi \times 0.1$ MHz. As shown in Fig. 2, we consider the process of a photon transfer from the electrical resonator to the mechanical one, and the efficiency is obtained to be larger than 90\%. The figure also shows that, during the transfer process, the excited qubit levels $|1\rangle$ and $|2\rangle$ are almost unpopulated, and thus verifies our adiabatic elimination of these states.


To sum up, when replacing the capacitor in circuit electromechanics by a superconducting flux qubit, we show that the coupling among the qubit and the two resonators can induce effective electromechanical coupling which attains the strong coupling regime at the single photon level with feasible experimental parameters.   Therefore, our work constitutes the first step towards studying nonlinear effect in circuit electromechanics.

\bigskip
\emph{Note added in the proof}: During the review process of this paper, a relevant experiment\cite{36} appears, which demonstrates the conversion of two microwave photons with different colors using two transmission line resonators.

\bigskip
This work was supported by the NFRPC (No. 2013CB921804) and the PCSIRT  (No. IRT1243).


\begin{thebibliography}{99} {\small

\bibitem{om} M. Aspelmeyer, T. J. Kippenberg, and F. Marquardt,  Rev. Mod. Phys. \textbf{86}, 1391 (2014).


\bibitem{em1} F. Marquardt, J. P. Chen, A. A. Clerk, and S. M. Girvin,
Phys. Rev. Lett. {\bf 99}, 93902 (2007).

\bibitem{linear1} J. D. Teufel, T. Donner, D. Li, J. W. Harlow, M. S. Allman, K. Cicak, A. J. Sirois, J. D. Whittaker, K. W. Lehnert, and R. W. Simmonds,
    Nature \textbf{475}, 359 (2011).

\bibitem{strong} T. A. Palomaki, J. D. Teufel, R. W. Simmonds, and K. W. Lehnert, Science {\bf 342}, 710 (2013).

\bibitem{em2}  T. A. Palomaki, J. W. Harlow, J. D. Teufel, R. W. Simmonds and K. W. Lehnert, Nature    \textbf{495}, 210 (2013).


\bibitem{em3} T. Bagci, A. Simonsen, S. Schmid, L. G. Villanueva, E. Zeuthen, J. Appel, J. M. Taylor, A. S{\o}rensen, K. Usami, A. Schliesser, and E. S. Polzik,
    Nature \textbf{507}, 81 (2014).

\bibitem{linear2} J. Chan, T. P. M. Alegre, A. H. Safavi-Naeini, J. T. Hill, A. Krause, S. Groblacher, M. Aspelmeyer, and O. Painter,
   Nature  \textbf{478}, 89 (2011).


\bibitem{enhance1} T. T. Heikkil\"{a}, F. Massel, J. Tuorila, R. Khan, and M. A. Sillanp\"{a}\"{a}, Phys. Rev. Lett. {\bf 112}, 203603 (2014).

\bibitem{en1}     J. R. Johansson, G. Johansson, and F. Nori, Phys. Rev. A {\bf 90},
053833 (2014).

\bibitem{en2} A. J. Rimberg, M. P. Blencowe, A. D. Armour, and P. D.
Nation, New J. Phys. {\bf 16}, 055008 (2014).

\bibitem{enhance2} J.-M. Pirkkalainen, S. U. Cho, F. Massel, J. Tuorila, T. T. Heikkil\"{a}, P. J. Hakonen, and M. A. Sillanp\"{a}\"{a}, Nat. Commun.  {\bf 6}, 6981 (2015).

\bibitem{enhance3} G. Via, G. Kirchmair, and O. Romero-Isart,
    Phys. Rev. Lett. {\bf 114}, 143602 (2015).


\bibitem{en3} E.-j. Kim, J. R. Johansson, and F. Nori,
Phys. Rev. A {\bf 91}, 033835 (2015).

\bibitem{mode1} A. Xuereb, C. Genes, and A. Dantan, Phys. Rev. Lett. {\bf 109},
223601 (2012); Phys. Rev. A {\bf 88}, 053803 (2013).

\bibitem{mode2} X.-Y. L\"{u}, Y. Wu, J. R. Johansson, H. Jing, J. Zhang, and F. Nori, Phys. Rev. Lett. {\bf 114}, 093602 (2015).

\bibitem{mode3} P.-B. Li, S.-Y. Gao, H.-R. Li, and F.-L. Li, arXiv: 1503.02393.


\bibitem{t1} I. Buluta, F. Nori,
Science  {\bf 326}, 108 (2009).


\bibitem{t2} Z.-L. Xiang, S. Ashhab, J. Q. You, and F. Nori,
Rev. Mod. Phys.  {\bf 85}, 623 (2013).

\bibitem{t3} I. M. Georgescu, S. Ashhab, and F. Nori,
Rev. Mod. Phys.  {\bf 86}, 153 (2014).

\bibitem{cqed} R. J. Schoelkopf and S. M. Girvin, Nature {\bf 451}, 664 (2008);
M. H. Devoret and R. J. Schoelkopf, Science {\bf 339}, 1169 (2013).

\bibitem{3l1} Y.-X. Liu, J. Q. You, L. F. Wei, C. P. Sun, and F. Nori,
Phys. Rev. Lett. {\bf 95}, 087001 (2005).

\bibitem{3l} J. Q. You and F. Nori, Nature {\bf 474}, 589 (2011).


\bibitem{3lexp} F. Deppe, M. Mariantoni, E. P. Menzel, A. Marx, S. Saito, K. Kakuyanagi, H. Tanaka, T. Meno, K. Semba, H. Takayanagi, E. Solano, and R. Gross, 
    Nat. Phys. {\bf 4}, 686 (2008).


\bibitem{c1} J. Johansson, S. Saito, T. Meno, H. Nakano, M. Ueda, K. Semba,
and H. Takayanagi, Phys. Rev. Lett. {\bf 96}, 127006 (2006).

\bibitem{fluxcavity} A. A. Abdumalikov, O. Astafiev, Y. Nakamura, Y. A. Pashkin,
and J. S. Tsai, Phys. Rev. B {\bf 78}, 180502(R) (2008).


\bibitem{c2} A. Fedorov, A. K. Feofanov, P. Macha, P. Forn-D¨ªaz, C. J. P. M. Harmans, and J. E. Mooij, Phys. Rev. Lett. {\bf 105}, 060503 (2010).


\bibitem{c3} M. Stern, G. Catelani, Y. Kubo, C. Grezes, A. Bienfait, D. Vion, D. Esteve, and P. Bertet,
    Phys. Rev. Lett. {\bf 113}, 123601 (2014)


\bibitem{m1}  A. D. O'Connell, M. Hofheinz, M. Ansmann, R. C. Bialczak, M. Lenander, E. Lucero, M. Neeley, D. Sank, H. Wang, M. Weides, J. Wenner, J. M. Martinis, and A. N. Cleland,
    Nature {\bf 464}, 697 (2010).


\bibitem{tune1} J. Q. You, Y.-x. Liu, C. P. Sun, and F. Nori,
Phys. Rev. B {\bf 75}, 104516 (2007);
J. Q. You, Y.-x. Liu, and F. Nori,
Phys. Rev. Lett. {\bf 100}, 047001 (2008).


\bibitem{tune2}  F. G. Paauw, A. Fedorov, C. J. P. M Harmans, and J. E. Mooij,
Phys. Rev. Lett. {\bf 102}, 090501 (2009).


\bibitem{ultrastrong} J. Bourassa, J. M. Gambetta, A. A. Abdumalikov, Jr., O. Astafiev, Y. Nakamura, and A. Blais, Phys. Rev. A {\bf 80}, 032109 (2009).


\bibitem{t4} S. Ashhab and F. Nori,
Phys. Rev. A  {\bf 81}, 042311 (2010).

\bibitem{eliminate} E. Brion, L. H. Pedersen, and K. M{\o}lmer, J. Phys. A {\bf 40},
1033 (2007).


\bibitem{nm1} A. Gaidarzhy, M. Imboden, P. Mohanty, J. Rankin, and B. W. Sheldon,
Appl. Phys. Lett. {\bf 91}, 203503 (2007).

\bibitem{nm2} Ya. S. Greenberg, Yu. A. Pashkin, and E. Il'ichev, Phys.-Usp. {\bf 55}, 382 (2012).
    

\bibitem{36} Z. H. Peng, Y.-x. Liu, J. T. Peltonen, T. Yamamoto, J. S. Tsai, and O. Astafiev, e-print arXiv: 1505.02858.

}
\end{thebibliography}
\end{document}